**Sharp Spectral Variations of the Ultrafast Transient Light Extinction by Bimetallic Nanoparticles in the Near-UV**

*Tadele Orbula Otomalo, Lorenzo Di Mario, Cyrille Hamon, Doru Constantin, Khanh-Van Do, Patrick O'Keeffe, Daniele Catone, Alessandra Paladini, and Bruno Palpant\**

Dr. T. Orbula Otomalo, Dr. K.-V. Do, Prof. B. Palpant
Université Paris-Saclay, CNRS, ENS Paris-Saclay, CentraleSupélec, LuMIn, 3 rue Joliot Curie, 91190 Gif-sur-Yvette, France
E-mail: bruno.palpant@universite-paris-saclay.fr

Dr. L. Di Mario, Dr. D. Catone
Istituto di Struttura della Materia - CNR (ISM-CNR), Division of Ultrafast Processes in Materials (FLASHit), Via del Fosso del Cavaliere 100, 00133 Rome, Italy

Dr. C. Hamon, Dr. D. Constantin
Université Paris-Saclay, CNRS, LPS, rue Nicolas Appert, bâtiment 510, 91400 Orsay, France

Dr. P. O'Keeffe, Dr. A. Paladini
Istituto di Struttura della Materia - CNR (ISM-CNR), Division of Ultrafast Processes in Materials (FLASHit), 00016 Monterotondo Scalo, Italy

Noble metal nanoparticles exhibit localized plasmon resonance modes that span the visible and near-infrared spectral ranges and have many applications. Modifying the size, shape, and composition of the nanoparticles changes the number of modes and their properties. The characteristics of these modes are transiently affected when illuminating the nano-objects with ultrashort laser pulses. Here, we synthesize core-shell gold-silver nanocuboids and measure their spectral signature in the stationary and ultrafast transient regimes. Their dipolar transverse mode vanishes with increasing Ag-shell thickness, while higher-order modes grow in the near-ultraviolet range where no plasmon resonance can be generated with single noble metal nanoparticles. These higher-energy modes are associated with sharp spectral variations of the ultrafast transient light extinction by the bimetallic nanocuboids. By carrying out a theoretical investigation, we break down the different contributions to this response and provide a physical interpretation of its spectral profile. The transient optical signal is then shown to reveal resonance modes hidden in the stationary regime spectra.

## 1. Introduction

The localized surface plasmon resonance (LSPR) of metal nanoparticles (NPs) can be tailored by controlling their morphology and composition[1–5] so that multiple modes may appear. Rod-like metal nanostructures that possess two LSPRs corresponding to field polarization along their longitudinal axis (LgSPR) and transverse axis (TrSPR) are an iconic example. Owing to the tunability of their LgSPR within the visible and near-infrared spectral ranges[6] as well as their biocompatibility, gold nanorods (AuNRs) have frequently appeared in biochemical applications[7], plasmon-enhanced spectroscopies[8,9] and photothermal therapy[7,10,11] developments. Moreover, the field intensity enhancement factor of silver nano-objects can be 6.8 times greater than that of their gold counterparts.[4] However, preparing silver nanorods of controlled morphology in water is still challenging. Fortunately, the outstanding optical properties of silver NPs can be combined with the convenience of gold NP synthesis resulting in AuNR-core and Ag-shell hybrid bimetallic nanostructures (AuNR@Ag).[3,4,12–16] The Ag deposits preferentially along the transverse dimension of the AuNR, leading to cuboid-shaped AuNR@Ag NPs. Their optical properties can be tuned from pure AuNRs'[6] to pure Ag cuboids'[17] by tailoring their equivalent Ag:Au molar ratio. These NPs have already been successfully applied in numerous disciplines in the stationary regime, some of which are Raman signal enhancement[18,19], biomedicine[20,21], sensing[22], and colorimetric assays of visual readout.[23] Furthermore, the sensitivity of the acoustic vibration modes (phonon dynamics) of AuNRs[24] to the amount of material deposited onto their surface has been a promising prospect for mass sensing applications, i.e., as a nano-balance.[25–28]

The transient optical response of plasmonic NPs originates from the series of fast energy exchanges they undergo upon their interaction with a laser pulse. Despite the variety of spectral and temporal features observed in such nanostructures, a limited number of works have been devoted to the transient electron dynamics of anisotropic Au-Ag bimetallic NPs, all of them being mainly experimental. Zarick *et al.* studied the electron dynamics in both cubic and pyramidal Au-core Ag-shell bimetallic nanostructures.[29] Yu *et al.* demonstrated that some of the resonance modes of the AuNR@Ag samples of small Ag-shell thickness (2 and 4 nm) that are hidden in the stationary regime spectra can be uncovered by ultrafast transient spectroscopic experiments.[15] Unfortunately, these studies were limited to a restricted spectral domain in the visible range. Furthermore, the investigation of the transient optical response of other kinds of bimetallic nano-objects has been reported in the literature as, for instance, in the case of Au@Ag and Au@Ag@Pt nanospheres[30], Au@Pd nanorods[25], Au@Pt nanospheres[31,32], Au-Pt hybrid NPs.[33]

The present work is devoted to the experimental and theoretical study of the stationary and transient optical responses of AuNR@Ag NPs from the near-ultraviolet to the red end of the visible range (300 – 800 nm). Their dependence on the Ag shell thickness will be mainly investigated. The AuNR@Ag exhibit multiple plasmonic resonance modes extremely sensitive to the amount of Ag deposited on the AuNR core. The main contributions to these modes will be discussed in the light of theoretical calculations. Then, we will describe the results of broadband ultrafast transient spectroscopy experiments. Each resonance mode identified in the stationary spectra is associated with a typical plasmon bleaching pattern in the transient optical response. However, the transient spectral signature of the thick Ag-shell NPs exhibits an additional bleaching pattern relative to the number of stationary LSPR modes observed. Furthermore, the multipolar transverse modes in the near-ultraviolet range result in a series of robust, steep, and spectrally narrow extinction variations. In the final discussion, the transient spectral signature's main contributions will be addressed, and the origin of the unexpected additional feature will be identified thanks to simulation results.

## 2. Optical Properties of the Nanoparticles in the Stationary Regime

The samples are labeled by their equivalent Ag:Au molar ratio, indicated by "eq". Five samples are considered, from pure AuNRs (eq=0) to those coated with a thick silver shell (eq=8). As shown in the TEM images (Figure S3, Supporting Information), the NPs grow into a cuboid shape as the Ag shell thickness increases.[2,3] The average dimensions of the NPs (Table S1, Supporting Information) determined by transmission electron microscopy (TEM) analysis show that in this synthesis, Ag grows preferentially along the transverse dimension, so the initial rod-like NPs take on a cuboid shape. In addition, as the Ag shell thickness increases, the edges and corners become sharper.[2,3] The extinction spectra of the colloidal samples in the stationary regime are presented in **Figure 1**(a). The pure AuNR dispersion exhibits two resonance peaks that correspond to the TrSPR and LgSPR modes' excitation at 515 nm and 680 nm, which will be further labeled by $T_0$ and $L_0$, respectively. The deposition of Ag leads to the appearance of additional peaks in the low-wavelength part of the spectrum. In the spectral range under investigation, results reported in the literature for similar bimetallic NP samples[2,4,12,34] exhibit four dominant resonance peaks. This is also the case here for the samples with the thinnest (0.8eq) and thickest (8eq) Ag shell. However, the extinction spectrum of the AuNR@Ag 2eq sample exhibits six peaks while that of the AuNR@Ag 4eq sample contains five peaks. Therefore, it is necessary to use new labeling as compared to previous papers to identify each peak. The mode labels are indicated in Figure 1(a) for all AuNR@Ag samples. As will be explained below, the most red-lying one (labeled by L) corresponds to a LgSPR mode while the others (labeled by T) are all TrSPR modes; $L_0$ and $T_0$ represent the longitudinal and transverse dipolar plasmon modes, respectively, while $T_1$, $T_2$, $T_3$ and possibly $T_4$ denote the higher-order transverse modes. The spectral evolution of the modes as a function of the Ag:Au molar ratio is reported in Figure S10 of the Supporting Information.

As depicted in Figure 1(a), the relative oscillator strength of the transverse $T_1$, $T_2$, and $T_3$ modes increases with the Ag shell thickness (the case of $T_4$ is difficult to discuss). The number of resonance modes appearing within our spectral range also evolves with the amount of Ag. It can be seen that the growth of Ag onto the AuNRs results in a blue shift of both the initial $L_0$ and $T_0$ resonances and a narrowing of the $L_0$ mode, called "plasmonic focusing"[35], compared to the LgSPR of the bare AuNRs [see Figure 1(a)]. In addition to this blue shift, one can notice the decrease in the relative magnitude of the $T_0$ mode as the Ag shell grows. It shows itself as a shoulder for the 2eq sample and is no longer visible for the larger amounts of Ag, only contributing to the asymmetric broadening of the $T_1$ mode of the 4eq sample. The higher-order plasmonic modes, $T_2$ and $T_3$, show a slight red shift as the Ag deposition increases.

The optical response of a representative NP for each sample was calculated using a numerical approach based on the boundary element method (BEM).[36] Details are given in the Supporting Information. The simulated extinction spectrum of a bare AuNR is reported in Figure 1(b) (black curve). The near-field map at the AuNR LgSPR and TrSPR modes can be found in Figure S5, Supporting Information. Figure 1(b) also displays the simulated extinction spectra of the bimetallic AuNR@Ag NPs, to be compared with the experimental ones [Figure 1(a)]. All the dominant resonances observed in the experimental spectra are also present in the calculated ones. As for the bare AuNRs, we have to keep in mind that the real samples exhibit a given NP size and shape distribution, explaining why the mode bands are broader and less peaked than the simulated spectra, which were calculated using a single NP morphology. The modeled data for the hybrid NPs present multiple small shoulders in the blue and red spectral regions of the $T_2$ and $T_3$ resonances, respectively. This can be seen readily from the plots in Figure 1(b) for the thick Ag-shell NPs. This irregularity of the higher photon energy resonances of the modeled data can be attributed to the influence of multimodal resonances originating from the sharpness of corners and edges (see Supporting Information § 3.1 for further comments). We provide a detailed description of the modes' physical origin in the Supporting Information, § 3.3.

## 3. Ultrafast Transient Optical Response

### 3.1. Experimental Investigation

We now focus on the spectrally-dependent transient optical response of monometallic AuNRs and bimetallic AuNR@Ag samples (depending on the Ag-shell thickness) after being irradiated by a femtosecond laser pulse of fluence 446 µJ cm$^{-2}$ and at a wavelength of 380 nm. The latter is chosen to excite the interband transitions of gold and/or the transverse T$_2$ mode of the bimetallic NPs, which peaks within a narrow spectral range from one sample to the other (see Figure 1). The transient spectral profiles of the absorbance variation, $\Delta A$, for the five samples are displayed in **Figures 2–4**. The transient absorption (TA) spectra of the bare AuNRs (Figure 2) agree with the literature of colloidal AuNRs solutions and single isolated AuNRs.[37,38] The transient signature shows alternative induced extinction (positive features) and induced transparency (negative features) signals that are due to the modification of the dielectric function of the AuNR after pump pulse absorption.[38] These features result from the sum of the transient responses of the two LSPR modes associated with the AuNR, namely, TrSPR and LgSPR [see the black curve in Figure 1(a)]. *Plasmon bleaching* denotes the damping and more or less symmetrical broadening of the plasmon absorption band. It translates into a negative $\Delta A$ at the band peak and a positive $\Delta A$ at the band red and blue sides. Two typical plasmonic bleaching signals can be observed around the TrSPR and LgSPR [see Figure 2(b) and (c)]. The central positive peak results from the superimposition of two positive signals contributed by the TA responses from the two plasmon modes of the AuNR.

It has been previously demonstrated that the relaxation dynamics varies with probe photon energy, which results in the spectral shift of the observed local maxima and/or minima.[38,39] This behavior can be seen readily in Figure 2(c), where the bleaching feature around the TrSPR blue-shifts with increasing delay time.

As expected, the TA signatures of AuNR@Ag 0.8eq [Figure 3(b) and (c)] and AuNR@Ag 2eq [Figure 3(f) and (g)] bimetallic samples, with the thinner Ag shells, present plasmonic bleaching features associated with the respective resonance modes identified in the stationary regime [Figure 3(a) and (e), respectively]. Moreover, for the thicker Ag shell samples AuNR@Ag 4eq [Figure 4(b) and (c)] and AuNR@Ag 8eq [Figure 4(f) and (g)], an unexpected weak negative peak can be observed between the bleaching features associated with their $T_1$ and $L_0$ resonances [see the green arrows in Figure 4(c) and (g)] where no stationary mode could be observed [Figure 4(a) and (e)]. The origin of this feature will be discussed and explained in section 4.3. Among the different investigations of the transient optical response of bimetallic NPs reported in the literature, only a few have discussed and analyzed the spectral profile of this response. Plasmon bleaching features were observed around the dipolar TrSPR and LgSPR modes of Au@Ag nanorod solutions with a wide shape distribution.[15] The same kind of transient spectral profile was revealed in bimetallic Au-Ag core-shell rounded nanocubes and nanopyramids.[29] Unfortunately, in all these works, the limited bandwidth of the white-light probe did not allow the observation of the higher-order modes lying at wavelengths lower than 450 nm. A plasmon bleaching pattern was also identified in the ultrafast transient response of Au nanospheres coated with a very thin layer of Pt-Au alloy, which exhibit only one plasmon mode (stemming from the Au core).[32]

Let us also mention that the transient optical response at large time delays (reported in the Supporting Information, § 6, for two samples) exhibits a recovery to the stationary properties on the timescale of hundreds of picoseconds, i.e., the transient spectral profile tends towards a succession of bleaching patterns centered on the different SPR modes. Therefore, some of the initial bleaching patterns get red-shifted along with the relaxation. Furthermore, the transient signals present oscillations with different frequencies as a function of time delay, which are associated with acoustic vibrations of the NPs.[24,25,27,40] The analysis of these oscillations is beyond the scope of this paper and will be discussed in a future communication.

## 3.2. Simulation of the NP Transient Optical Response

We have modeled in §2 the stationary optical response of the samples. This has allowed us to determine the theoretical morphological parameters (dimensions and rounding) of the most representative NPs in each sample, used now to model their transient response. Note that our goal is to analyze the overall transient spectral dynamics both for the monometallic and bimetallic samples rather than to reproduce the experimental data quantitatively. We will first calculate the metals' ultrafast transient dielectric function and use it to model the time-dependent extinction cross-sections of the NPs. For this, the pump laser's optical parameters are chosen to be identical to the experimental ones: pump wavelength 380 nm, excitation fluence 446 µJ cm$^{-2}$, beam waist radius $w = 200$ µm, and pulse duration $\Delta t = 50$ fs. Assuming a Gaussian pulse, the instantaneous laser power absorbed by each NP, $P_{abs}$, was calculated from their absorption cross-section, $\sigma_{abs}$, at 380 nm by the following relation:

$$P_{abs} = \frac{\sigma_{abs} I_0}{V}. \tag{1}$$

$V$ is the NP volume and $I_0 = \frac{2E}{\phi \sqrt{2\pi}\,\pi w^2}$ is the laser peak intensity, where $\phi = \frac{\Delta t}{\sqrt{ln256}}$ and $E$ is the excitation pulse energy.[41]

As we intend to simulate the transient signal within a time range spanning over several picoseconds while disregarding the initial athermal regime[38,42], the three-temperature model (3TM) has been selected to model the thermal dynamics inside and around the NPs. Details regarding the 3TM can be found in previous papers (see Reference [43] for instance). It accounts for the heat release from the NP to its environment (water in the present case). The electron temperature dependence of the electron heat capacity and the electron-phonon coupling constant account for the influence of d-band electrons in the density of states, as calculated by Lin *et al.*[44] The coupled nonlinear equations of the 3TM were solved via a numerical finite element method (FEM) optimized in a commercial software (COMSOL) as any analytical solving is excluded due to the complex morphology of the bare AuNR and AuNR@Ag NPs. A perfect contact at the interfaces and a purely diffusive thermal transport in the host medium were considered.[39] This provides us with the time evolution of the electron ($T_e$) and lattice ($T_l$) temperatures in the NPs and the topography of the host medium temperature.

The thermal dynamics modeling in and around the AuNR@Ag NPs was carried out by considering each bimetallic system as a single homogeneous cuboid NP. A more rigorous approach would have consisted in accounting for the thermal resistance at the Au/Ag interface and the difference in the electron-phonon (e-ph) coupling constants, heat capacities, and thermal conductivities of the two metals, which would have induced a discontinuity in the thermal dynamics at the boundary.[45] However, for the case of an Au/Ag interface, it has been shown by Yilbas[46] that the short-pulse induced dynamics of both $T_e$ and $T_l$ are smooth (i.e., without discontinuities at the interface). Therefore, the influence of the Au/Ag thermal resistance is negligible. This stems from the similarity between the thermal properties of Au and Ag (see Supporting Information, Table S3). This supports our simplified approach, which amounts to consider the bimetallic AuNR@Ag NP model as a homogeneous single cuboid for its thermodynamic properties. Let us underline that this simplification could not have been possible with other metal combinations, such as Au and Pt.[31–33]

In this way, the time evolution of both $T_e$ and $T_l$ inside the NPs have been determined.

These values have been used to calculate the transient variation of the dielectric functions of Au and Ag following the method already used previously for Au. Briefly, the respective electron distributions in Au and Ag are determined from the quasi-free electron gas temperature at each time step. Together with $T_e$ and $T_l$ it is then used to evaluate the electron-electron (e-e) and e-ph collision rates involved in the Drude-Sommerfeld model for the intraband susceptibility. The transient modification of the intraband susceptibility is added to the interband one, determined through Lindhard's theory and Rosei's model by considering a local parabolic band structure around the $L$ and $X$ points of the Brillouin zone for Au and at $L$ point for Ag. This assessment also relies on the electron distribution dynamics in both metals at each time step. Once the time evolution of the dielectric function is calculated, the subsequent time evolution of the optical properties of the NPs (extinction, scattering and absorption cross-sections) is determined using the BEM. The volume of the NPs is calculated according to their geometry. The AuNRs have a hemispherical-capped cylinder shape, while the AuNR@Ag bimetallic NPs are considered simple cuboids as explained above (the rounding is neglected for evaluating the NP volume).

The calculated transient extinction spectrum of a bare AuNR in water is shown in Figure 2(d) in the form of the difference between the transient extinction cross-section and its stationary value, $\Delta\sigma_{ext}(t)$. The simulated curves at different delay times after excitation exhibit two bleaching peaks, just like in the experimental spectra with the colloidal AuNRs in Figure 2(c). The magnitude of the calculated transient signal associated with the TrSPR is tiny compared to that of the LgSPR. The bleaching signal around the TrSPR blue shifts with the delay time, as in the experimental data [Figure 2(c)]. Let us mention that the analysis of the transient optical response of bare AuNRs has already been reported in the literature.[37,38] Figure 3(d), 3(h), 4(d) and 4(h) depict the calculated transient optical responses of the AuNR@Ag 0.8eq, AuNR@Ag 2eq, AuNR@Ag 4eq and AuNR@Ag 8eq NPs, respectively.

The data display the same number of dominant plasmonic bleaching peaks, as has been observed in their experimental counterpart [Figure 3 and 4, panels (c) and (g)]. For each sample, the spectral positions of the bleaching peaks of the calculated data agree very well with the experimental ones. While for the bare AuNR the experimental and simulated spectral signatures are similar over the whole spectral region of interest, their agreement for the AuNR@Ag samples is less obvious in the higher photon energy range. Multiple spikes can be observed in the simulated differential cross-section for the bimetallic samples. The spikes get stronger for the larger Ag-shell thickness NPs [see Figure 3 and 4, panels (b) and (e)]. As discussed before, this is due to the presence of multimodes in this spectral zone, which bundle together to form each peak of the stationary spectrum. Their contribution to the transient signal is amplified compared with their contribution to the stationary extinction spectrum, due to the strong and narrow differential signal sign change associated with each resonance mode. Besides, the oscillator strength weights of the positive peaks in this spectral zone are somehow overestimated compared to what can be seen in the experimental data. This overestimation has the same origin as the difference in the sharpness of the spectral features observed in the stationary regime [compare the curves in Figure 1(a) with their calculated counterparts in Figure 1(b)]. Again, the theoretical calculation considers only a single NP geometry: the experimental sample's shape and size distributions are not accounted for. These distributions result in the blurring of the spectral variations measured compared with the spikes in the simulated signal.

### 3.3. Discussion

We now analyze the samples' ultrafast transient optical response in light of the theoretical model described in § 3.2. **Figures 5 and 6** exhibit the pump-induced variation of the extinction, absorption, and scattering cross-sections at a delay time $t$ = 2.5 ps for the thinnest and the thickest Ag-shell NPs. Consistent with what is discussed in § 3.4 of the Supporting Information regarding the NP size effect in the formation of the spectra in the stationary regime, the main contribution to the transient spectral signature of the thin Ag-shell NP comes from the absorption process [compare the curves and their magnitudes in Figure 5(d)]. In addition, the optical response of the Ag shell in the spectral zone of the LgSPR is small. The $T_0$ mode bleaching peak of the AuNR@Ag 0.8eq NP is very close to that of its $L_0$ mode, and its oscillator strength has not been reduced much compared to the bleaching peak of the TrSPR of the bare AuNR [Figure 2(d)].

The role of absorption in dictating the LgPol mode fades and scattering takes over as the Ag shell thickness increases, namely in the case of AuNR@Ag 8eq (see Figure 6). As in the stationary regime analysis, both scattering and absorption cross-sections contribute (almost equally) to the formation of the overall transient spectral signature of the extinction cross-section. Absorption is mainly responsible for the features in the spectral range of the TrSPR modes, while scattering is dominant in the LgSPR range [compare the curves in Figure 6(d)]. Moreover, both contributions compete around the $L_0$ mode (opposite signs), whereas they cumulate in the transverse modes' range (at higher photon energy). This explains the increase of the high-energy transverse modes' relative contribution when increasing the Ag shell thickness, as observed in the experimental and simulated overall spectra.

Let us also notice that the transverse polarization (TrPol) component of the transient absorption [purple curve in Figure 6(b)] plays a crucial role in the formation of the fifth bleaching peak identified in the transient spectral signature of the thick Ag-shell NPs [see the green arrows in Figure 4, panels (d) and (h)]. This transient bleaching peak is formed between the bleaching peaks of the $L_0$ and $T_1$ modes, which have both been identified experimentally in the stationary regime [(see Figure 1(a)] and reproduced theoretically [Figure 1(b)]. It means, together with the signal's transverse absorption origin, that the fifth peak results from the pump-induced modulation of the dipolar $T_0$ mode; however, this mode is concealed in the stationary regime by the Ag shell's influence, as discussed in the Supporting Information § 3.4. So, in the transient regime, the $T_0$ mode has forced itself into appearance thanks to the differential nature of the response and the fact that the transient signal experiences sign changes within a narrow spectral band. This highlights that the influence of small spectral variations of the dielectric function cannot be readily observed in the stationary regime, but their influence can be amplified in the transient optical response. The latter can then be exploited as a sensitive tool to reveal resonance modes, despite its complex spectral signature.

Let us finally describe the physical origin of the dynamics of the plasmon modes briefly, by discussing the simulated data [Figs. 2(d), 3(d), 3(h), 4(d), and 4(h)] which refer to a single NP morphology. The transient extinction signal reveals that around the $L_0$ maximum wavelength (dashed vertical line), the differential extinction profile is the result of a symmetric contribution to $\Delta A$ (bleaching of the mode) and a small antisymmetric one (red-shift of the mode), the magnitude of which depends on the Ag/Au molar ratio. Furthermore, the balance between these effects evolves with the delay time. As already discussed in a previous paper[38], for bare AuNRs at very short delay times ($\lesssim$ 1 ps) the response is dominated by the contribution of interband transitions, while at longer delay times ($\gtrsim$ 3 ps) the intraband transition contribution dominates. The precise dynamics of this balance may depend on (i) the excitation energy, as the electron heat capacity is ruled by the electron temperature reached, and (ii) the energy value of the LgSPR, as the oscillating forces associated respectively with the intraband and the interband transitions are spectrally dependent.

The addition of a silver shell preserves the dipolar nature of $L_0$: electrons in Ag and Au oscillate in-phase (see § 3.2 to 3.4 of the Supporting Information). Nevertheless, both intraband and interband transitions in silver may now be involved in the short-time dynamics of the $L_0$ mode, in addition to the ones of gold. Indeed, due to the high pump photon energy ($\hbar\omega_{pump} = 3.26$ eV), the electron distribution is initially perturbed far from the Fermi level (the electron distribution is affected in a region of $\pm\hbar\omega_{pump}$ around the Fermi energy). The metal's dielectric function can then be modified by changes in interband transitions, even for probe photon energies lower than the stationary interband transition threshold, which is the case here for the $L_0$ mode of the bare AuNRs and the thinner Ag-shell AuNR@Ag. This was already pointed out for gold[47] and silver[48,49] nanoparticles. The Au and Ag intraband and interband contributions, as well as their relative weights, change with increasing the Ag/Au molar ratio due to (i) the modification of the relative volume fractions of Au and Ag, of course, (ii) the spectral shift of the $L_0$ mode, and (iii) the transition from the absorptive nature of extinction to a scattering nature, as thoroughly discussed above. However, it can be seen that the spectral profile of the differential extinction induced by the initial pump pulse is roughly preserved in all NPs investigated, that is, a bleaching profile with an additional small antisymmetric component accounting for a slight red-shift of the resonance mode.

The analysis of the dynamics of the successive higher energy modes is qualitatively similar to that of $L_0$, with an increased relative contribution of silver on getting closer to the interband transition threshold and a progressive displacement of the field from the core to the shell (see § 3.3 of the Supporting Information). It is nevertheless made difficult by the spectral overlap of the transient signal features associated with each mode. When considering now the real samples with possible shape and size distributions, some features observed are quenched, stretched and/or blurred as compared with their monodisperse distribution counterpart.[38] Therefore, the discussion of the complexity of the joint physical contributions to the actual dynamics of the transient signals deserves a full and specific investigation that goes beyond the scope of the present article, focused on the analysis of the transient spectral signatures.

**4. Conclusion**

AuNRs were synthesized by a seed-mediated process, and an Ag shell was epitaxially grown on top of them. The synthesis allows for the deposition of more Ag mass along the sides of the AuNRs than on their tips leading to cuboid-shaped AuNR@Ag NPs. The Ag-shell thickness-dependent optical response of the samples has been studied both in the stationary and transient regimes. Multiple plasmonic resonance modes have been observed in the stationary regime extinction spectra of the samples. These modes are susceptible to the amount of Ag deposited on the AuNR core. The spectral positions of the two dipolar modes are strongly influenced by the Ag-shell deposition (they blue-shift as the Ag-shell thickness increases) due to the influence of the optical properties of Ag on their Au counterpart. Broadband pump-probe spectroscopy was carried out to record the ultrafast transient optical response of the samples. The experimental data have been well reproduced theoretically using the 3TM. The calculations explain all features in the transient spectra, including the unexpected transient signal between the bleaching patterns of the $L_0$ and $T_1$ modes in the thick Ag-shell AuNR@Ag samples. This feature was assigned to a pure dipolar $T_0$ mode which evolves from the TrSPR dipolar mode of the bare AuNR with increasing Ag coating. While it becomes hidden in the stationary regime spectra it is revealed by the ultrafast differential transient signal.

The bimetallic core-shell nanoparticles have then been shown to exhibit a strong ultrafast optical response in the blue and near-UV domains, a spectral range which is not addressed with usual monometallic plasmonic nanostructures. The transient features associated with the transverse mirror-charge resonance modes can even be more intense than that associated with the longitudinal plasmon mode. Furthermore, the number of modes, and thus the transient optical signature, can be tuned by controlling the NP morphology. The steep and spectrally-narrow variations of the absorption and near-field in the near-UV may be exploited for different purposes, such as time-resolved sensing, photonic switching, ultrashort timescale control of fluorescence, near-field enhancement modulation for time-resolved sensors, multiphotonic electron emission and photoluminescence in new spectral ranges. While observed here in an ensemble of nanoparticles with size and shape distributions, this complex transient response may be even sharper with single nano-objects, as predicted by our calculations. Beyond, these hybrid samples can be assembled into supercrystals[3], allowing modifying or improving the stationary and ultrafast transient optical properties through couplings, as highly spatially-organized metal nanorods are attractive for their robust and polarization-dependent plasmon properties.

## 5. Experimental Section

*Nanoparticle synthesis*: The synthesis of the bimetallic NPs is carried out in two successive steps. First, Au nanorods of $47.6 \pm 4.7$ nm length and $15.8 \pm 3.3$ nm width are elaborated in water (see Supporting Information § 1.2 and Figure S1). A silver shell of controlled thickness is then grown onto the gold nanorod core according to recently published protocols (see Supporting Information, Figure S2).[2,3,16] Details about the synthesis procedure and the reactants used can be found in the Supporting Information.

*Transient absorption experiments:* Transient absorption (TA) spectroscopy measurements were carried out by using a pump-probe scheme. Our laser setup consists of a Ti:Sapphire mode-locked oscillator, a regenerative amplifier, and an optical parametric amplifier (OPA). The tunable OPA output was used as the optical pump, with a transverse size of about 400 μm on the sample, a pulse length of about 50 fs and a repetition rate of 1 kHz. The colloidal solution samples were set in 1-mm thick quartz cuvettes. Based on the solution concentration, the number of NPs illuminated by the laser beam is estimated to be $150 \times 10^{6}$ in a probed volume of 0.125 μL. Under these conditions, after pump excitation, a complete dissipation of the heating induced by the pump occurs before the subsequent pulse. A white light supercontinuum generated in a femtosecond transient absorption spectrometer of IB Photonics (FemtoFrame II) was used as the probe. The probe wavelengths ranged between 330 and 770 nm, while the pump-probe delay time lasted up to 500 ps, with an overall instrument response function (IRF) of about 50 fs. The measurements reported were performed using a pump laser wavelength of 380 nm and a fluence of 446 µJ cm$^{-2}$. Let us mention that the variation of absorbance ($\Delta A$) measured as a function of both the probe wavelength and the delay time between pump and probe pulses results from the pump-induced variation of optical extinction within the sample. Details on the pump-probe experiment setup can be found in previous papers.[50,51]

**Supporting Information**

Supporting Information is available from the Wiley Online Library or from the author.

**Acknowledgements**

Claire Goldmann is acknowledged for technical support. T.O.O. thanks the "Groupement de Recherche Or-Nano" (GDR CNRS n° 2002) for the awarding of a research internship grant.

**Conflict of Interest**

The authors declare no conflict of interest.

Received: ((will be filled in by the editorial staff))

Revised: ((will be filled in by the editorial staff))

Published online: ((will be filled in by the editorial staff))

**Figures**

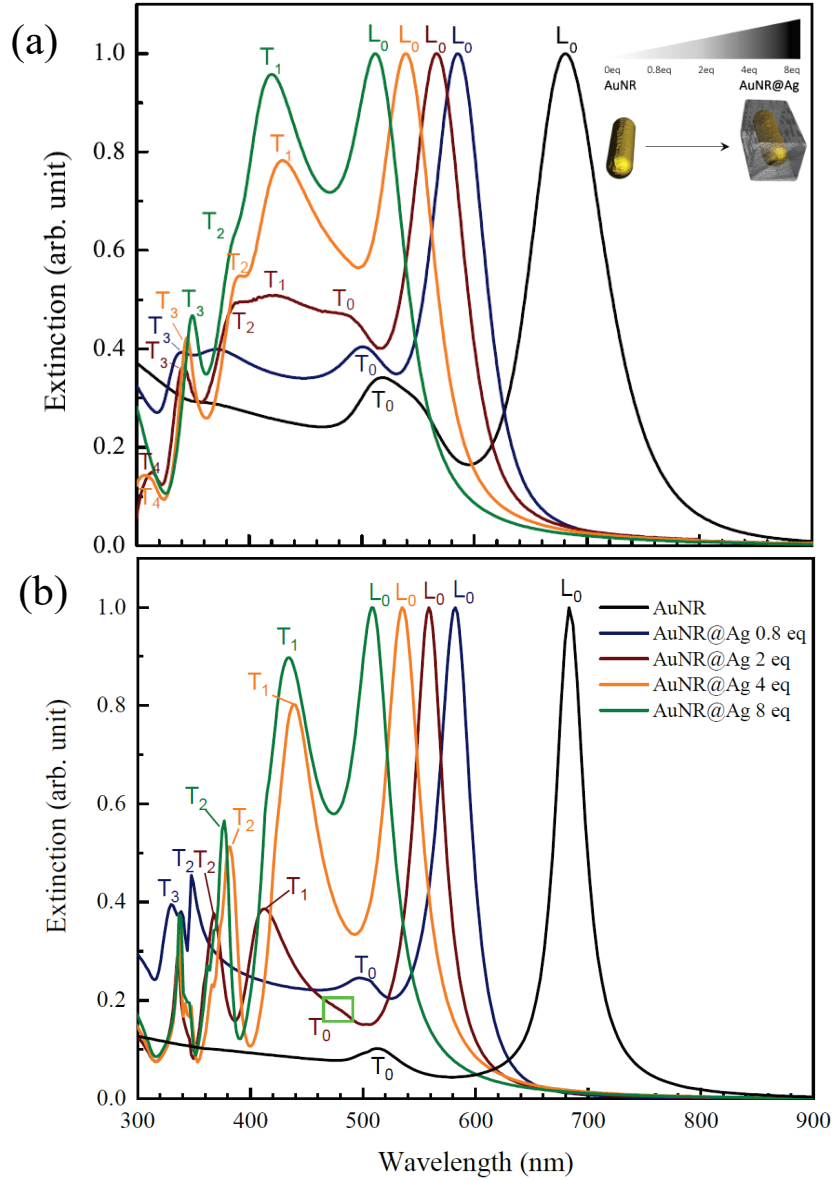

**Figure 1.** (a) Experimental stationary regime extinction spectra of the water solutions of uncoated AuNRs (black) and of the AuNR@Ag NPs. The data are normalized to their respective maximum values at the $L_0$ mode. The L and T labels represent the longitudinal and transverse modes, respectively. Subscripts of L and T denote the mode types that are present for each field polarization: subscript 0 denotes a dipolar mode while subscripts 1, 2, 3 and 4 represent higher-order modes excited in the transverse polarization. The inset depicts the Ag overgrowth process. (b) Extinction spectra of the representative AuNR@Ag NPs for each sample as calculated by using the BEM. All curves are normalized to their respective maximum value. The green square for 2eq highlights the presence of the weak $T_0$ mode.

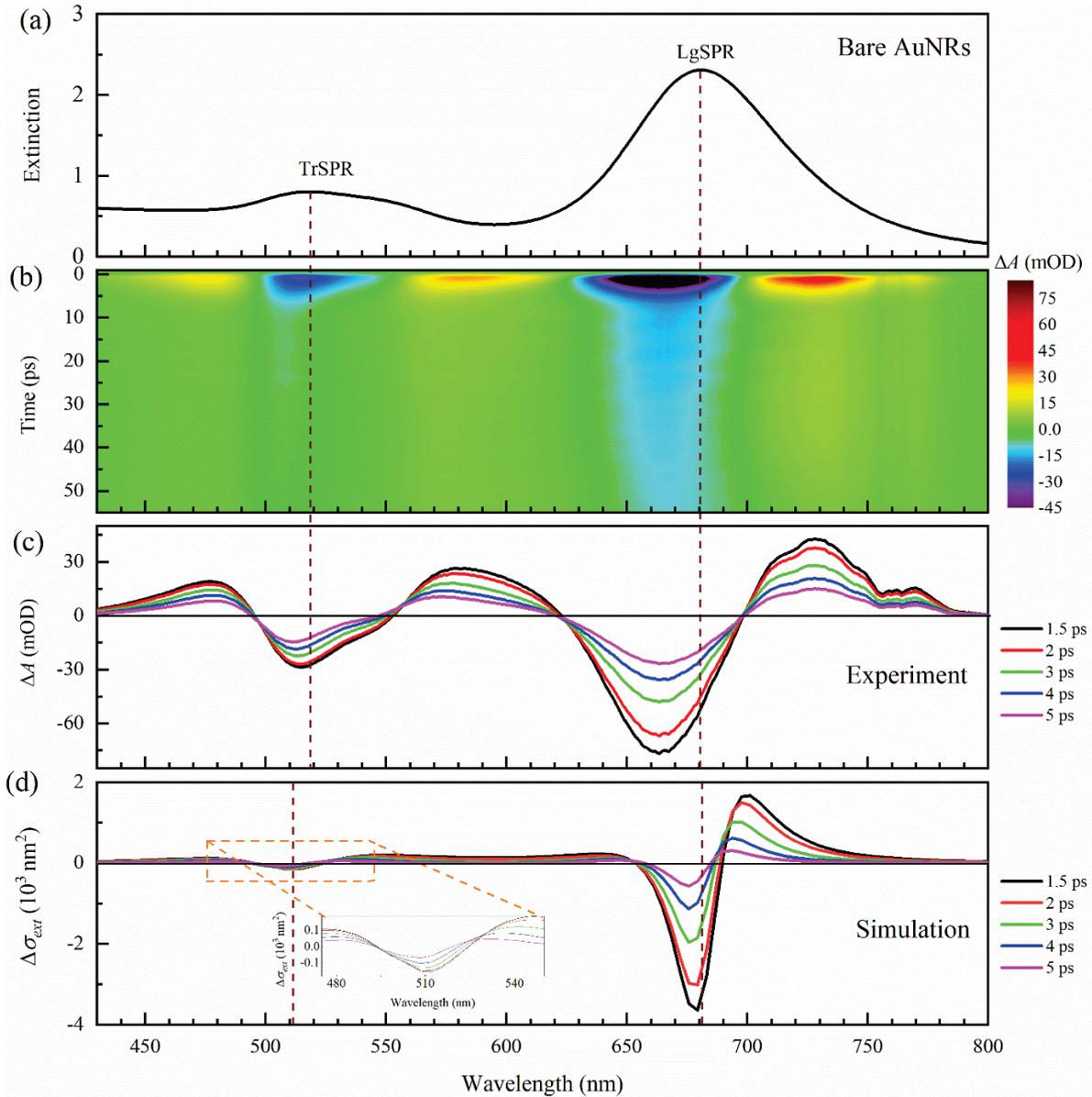

**Figure 2.** (a) Extinction of the AuNR solution sample acquired by UV-VIS spectrophotometry exhibiting the TrSPR and LgSPR bands. (b) False color map of the time and spectral dependence of the absorbance variation Δ*A* (in mOD) of the AuNRs measured via pump-probe spectroscopy. (c) Spectral dependence of the differential absorbance at different delay times as extracted from (b). (d) Simulated transient optical response (deviation of the extinction cross-section from its stationary value) of a bare AuNR with AR = 2.9 at different delay times. The inset shows the dynamics around the TrSPR mode. The vertical dashed lines indicate the experimental [panels (a)-(c)] or calculated [panel (d)] spectral position of the two plasmon resonance modes of the AuNRs in the stationary regime.

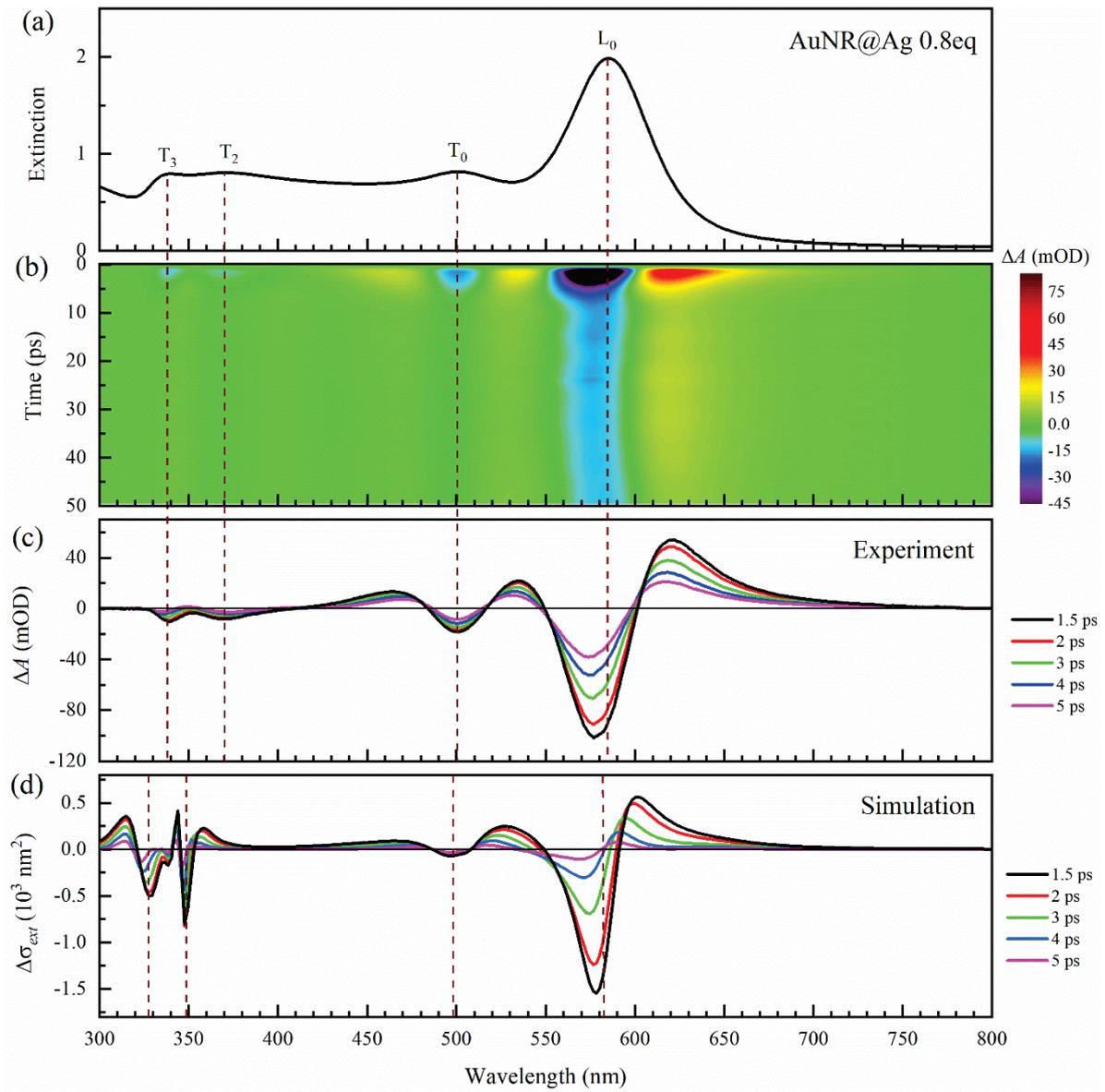

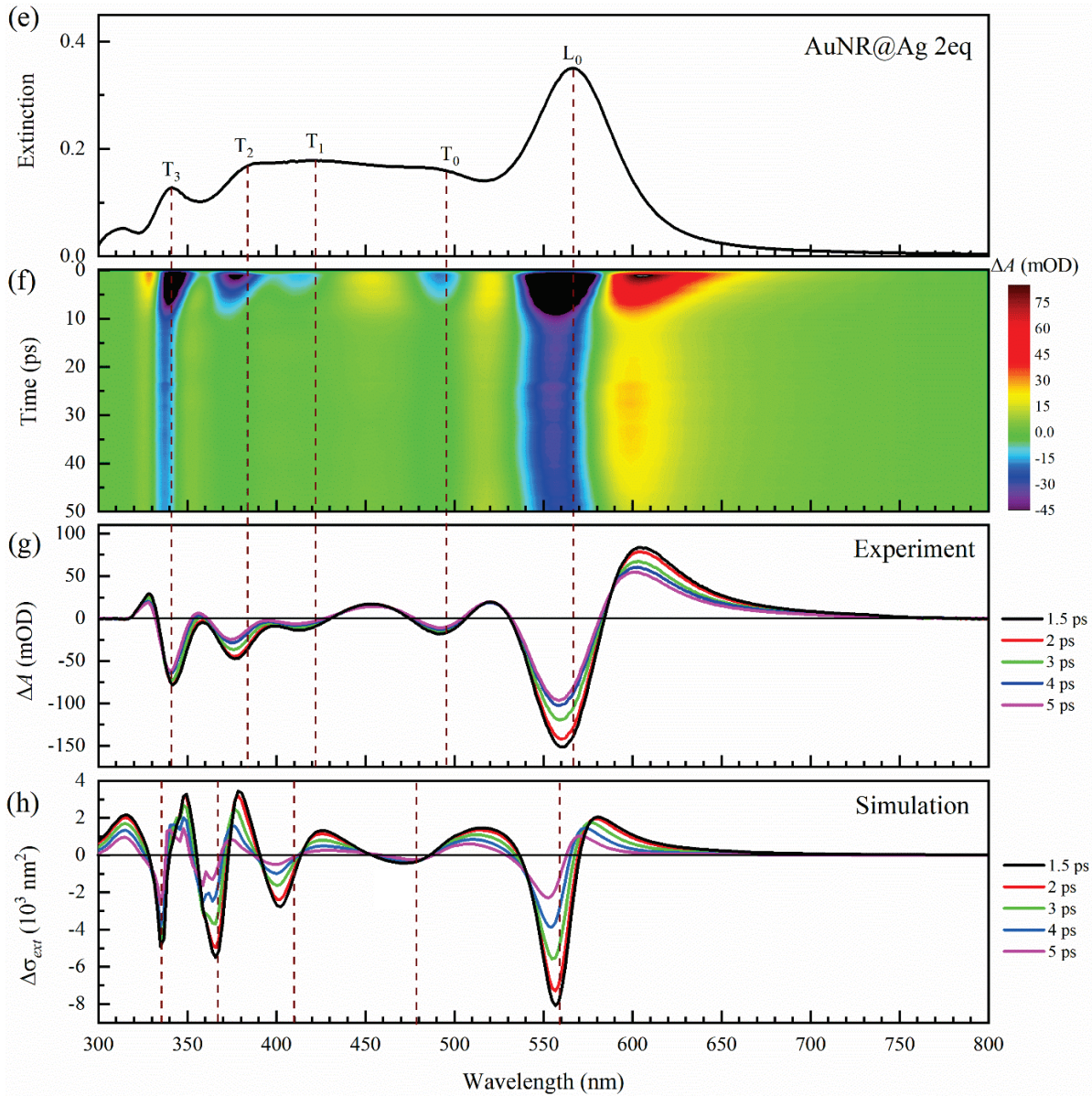

**Figure 3.** (a) and (e): Experimental extinction of the AuNR@Ag solution samples with thin Ag shell, 0.8eq (a) and 2eq (e) acquired by UV-VIS spectrophotometry. (b) and (f): False color map of the time and spectral dependence of the absorbance variation (in mOD) of the 0.8eq and 2eq samples, respectively, measured by pump-probe spectroscopy. (c) and (g): Spectral dependence of the differential absorbance at different delay times as extracted from (b) and (f), respectively. (d) and (h): Simulated transient optical response (deviation of the extinction cross-section from its stationary value) of the representative AuNR@Ag NPs with thin Ag shell, 0.8eq (d) and 2eq (h) at different delay times. The vertical dashed lines indicate the experimental [panels (a)-(c) and (e)-(g)] or calculated [panels (d) and (h)] spectral position of the main plasmon resonance modes.

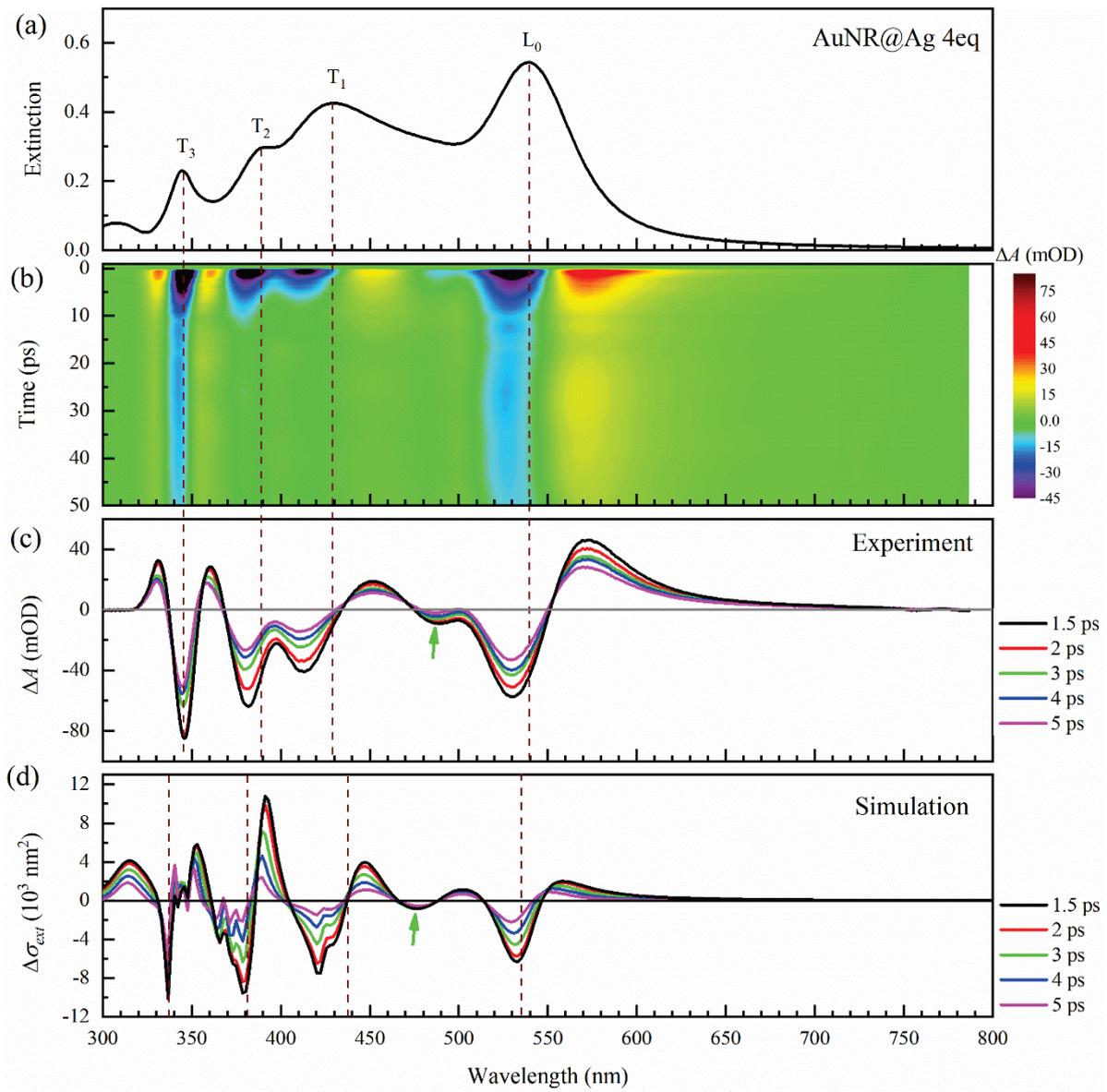

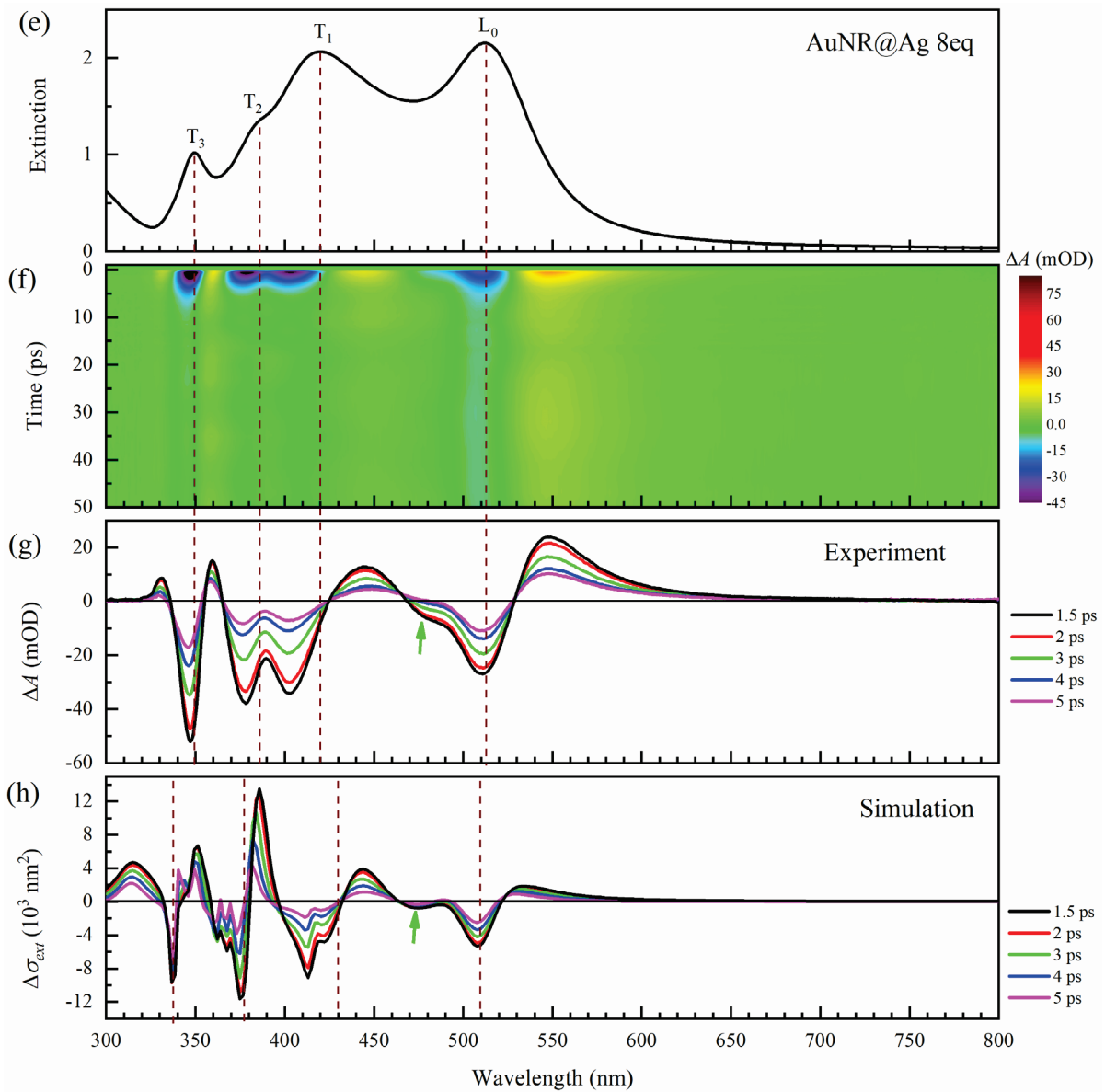

**Figure 4.** Same as Figure 3 but for the thick Ag shell samples AuNR@Ag 4eq [(a) to (d)] and AuNR@Ag 8eq [(e) to (g)]. The green arrow in (c) and (g) indicates the fifth bleaching peak associated with the weak $T_0$ mode which is hidden in the red wing of the broad $T_1$ band in (a) and (e) (see the discussion below).

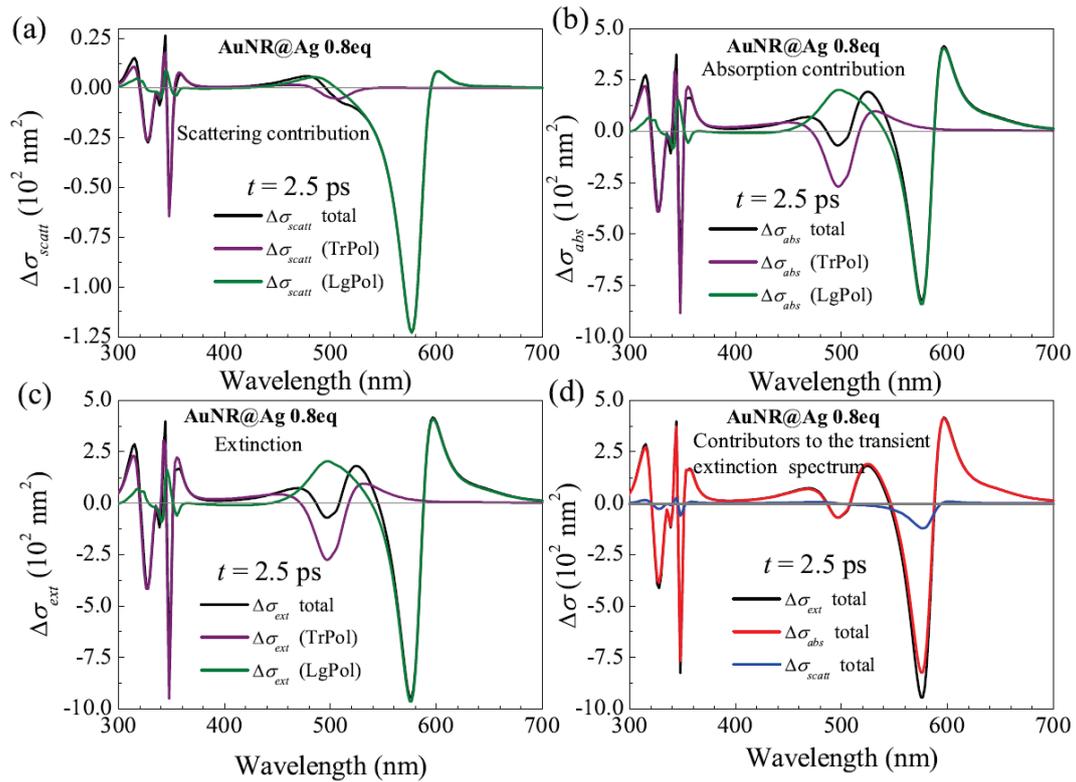

**Figure 5.** Contributions to the transient extinction cross-section of the AuNR@Ag 0.8eq (thin Ag-shell) sample at 2.5 ps delay. The scattering [black curve in (a)] and absorption [black curve in (b)] cross-section contributions to the transient extinction [black curve in (c)] spectral signature are also shown. The purple and olive colored curves in (a), (b) and (c) correspond to the transverse- (TrPol) and longitudinal-polarization (LgPol) excitation components of the corresponding optical properties. The blue and red colored curves in panel (d) are the scattering and absorption contributions to the total extinction (black curve).

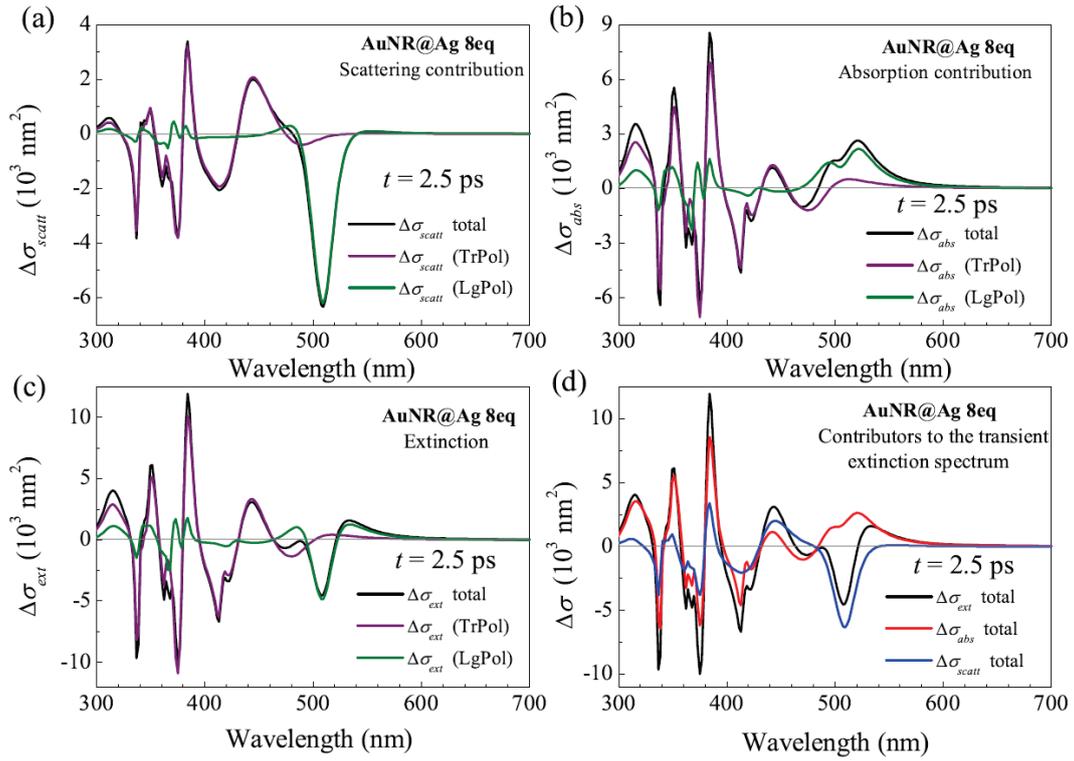

**Figure 6.** Same as Figure 5 but for AuNR@Ag 8eq (thick Ag shell).